\documentclass[letterpaper]{article} 
\usepackage{aaai24}  
\usepackage{times}  
\usepackage{helvet}  
\usepackage{courier}  
\usepackage[hyphens]{url}  
\usepackage{graphicx} 
\usepackage{natbib}  
\usepackage{caption} 
\usepackage{algorithm}
\usepackage{algorithmic}
\usepackage{xcolor}

\usepackage{newfloat}
\usepackage{listings}
\DeclareCaptionStyle{ruled}{labelfont=normalfont,labelsep=colon,strut=off} 
\floatstyle{ruled}
\newfloat{listing}{tb}{lst}{}
\floatname{listing}{Listing}
\title{DRGame: Diversified Recommendation for Multi-category Video Games \\ with Balanced Implicit Preferences}

\author{
   Kangzhe Liu\textsuperscript{\rm 1},
   Jianghong Ma\textsuperscript{\rm 1}\thanks{Corresponding author.},
   Shanshan Feng\textsuperscript{\rm 2},
   Haijun Zhang\textsuperscript{\rm 1},
   Zhao Zhang\textsuperscript{\rm 3}\\
}
\affiliations{
   \textsuperscript{\rm 1}
School of Computer Science and Technology, Harbin Institute of Technology, Shenzhen\\
\textsuperscript{\rm 2}
   Centre for Frontier AI Research, Agency for Science, Technology and Research, Singapore\\
   \textsuperscript{\rm 3}
School of Computer Science and Information Engineering, Hefei University of Technology, Hefei\\
kangzheliu@foxmail.com, majianghong@hit.edu.cn, victor\_fengss@foxmail.com, hjzhang@hit.edu.cn, cszzhang@gmail.com
}

\usepackage{bibentry}

\usepackage{amssymb}
\usepackage{amsmath}
\usepackage{subfigure}
\usepackage{booktabs}
\usepackage{multirow}

\begin{document}

\maketitle

\begin{abstract}
The growing popularity of subscription services in video game consumption has emphasized the importance of offering diversified recommendations. Providing users with a diverse range of games is essential for ensuring continued engagement and fostering long-term subscriptions. However, existing recommendation models face challenges in effectively handling highly imbalanced implicit feedback in gaming interactions. Additionally, they struggle to take into account the distinctive characteristics of multiple categories and the latent user interests associated with these categories. In response to these challenges, we propose a novel framework, named DRGame, to obtain diversified recommendation. It is centered on multi-category video games, consisting of two {components}: Balance-driven Implicit Preferences Learning for data pre-processing and Clustering-based Diversified Recommendation {Module} for final prediction. The first module aims to achieve a balanced representation of implicit feedback in game time, thereby discovering a comprehensive view of player interests across different categories. The second module adopts category-aware representation learning to cluster and select players and games based on balanced implicit preferences, and then employs asymmetric neighbor aggregation to achieve diversified recommendations. Experimental results on a real-world dataset demonstrate the superiority of our proposed method over existing approaches in terms of game diversity recommendations.


\end{abstract}

\section{Introduction}

In recent years, there has been a noteworthy shift in the consumption pattern of video games, gradually transitioning from traditional individual game purchases towards subscription services like Microsoft's Xbox Game Pass (XGP)\footnote{https://www.xbox.com/en-US/xbox-game-pass}. These subscription services offer users the opportunity to 
access all games available on the platform at a monthly fee, placing greater emphasis on the content of games rather than individual game prices. As a consequence, the role of diversity in game recommendation becomes crucial, as it encourages users to explore various types of games,
thereby promoting sustained and prolonged subscriptions. However, existing research in video game recommendation has predominantly focused on optimizing for accuracy, neglecting the vital aspect of diversity. 
This oversight raises concerns about the echo chamber/filter bubble \cite{echo-chamber/filter-bubble} effect,
where users are confined to familiar game genres, leading to diminished motivation in continuing their subscription once they have exhausted the limited choices available to them.

\begin{figure}[t]
\centering
\includegraphics[width=1\columnwidth]{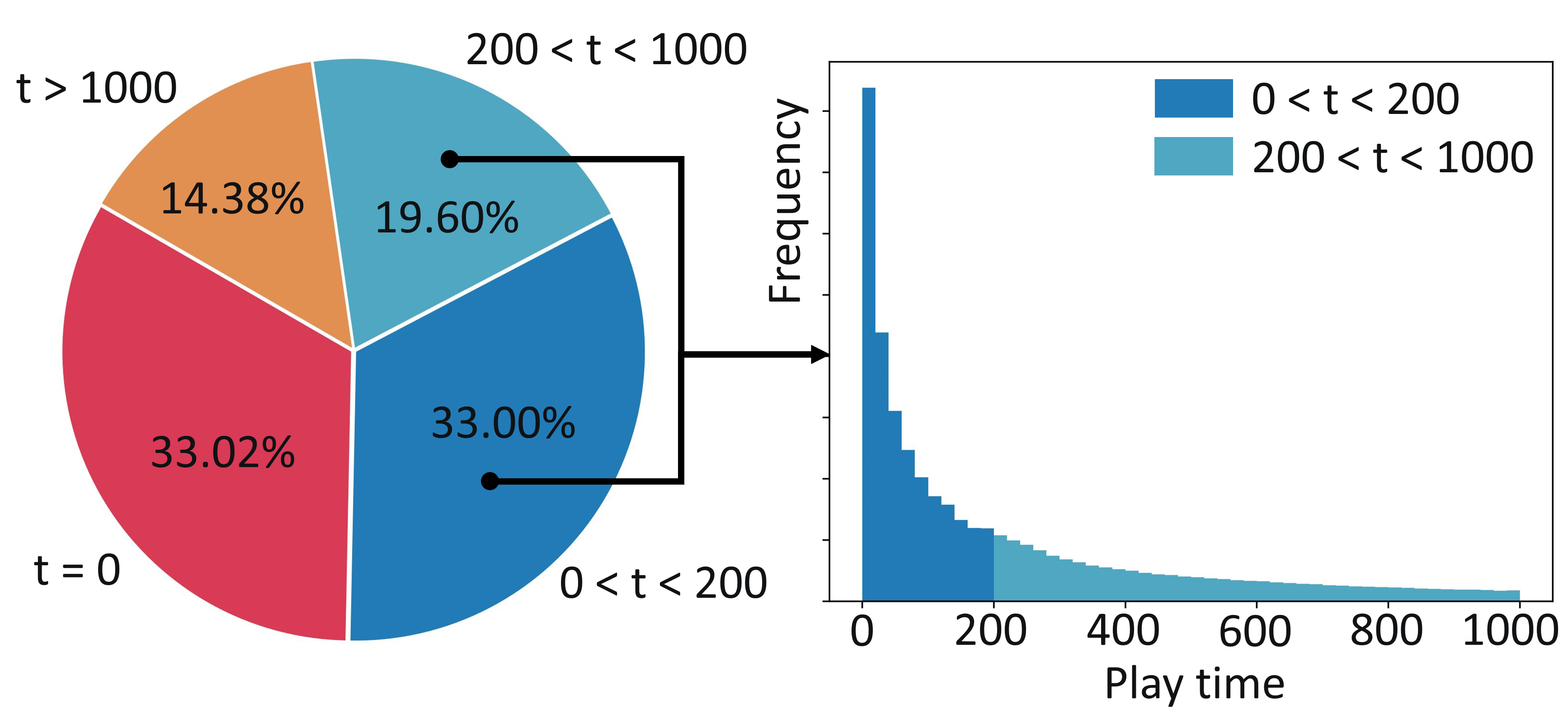} 
\caption{The distribution of playtime, which follows a power-law distribution and contains a substantial proportion of zero values.}
\label{fig_intro}  \vspace{-6mm}
\end{figure}

Diversified recommendation systems have been extensively studied in various areas, such as e-commerce, movies, music, and news, leveraging explicit or implicit feedback for effective user personalization \cite{ding2021hybrid}. {However, the domain of video games remains an unexplored frontier in this context. Additionally,} video game recommendation presents unique challenges due to the distinct nature of gaming interactions. Unlike other fields, explicit feedback in the form of reviews and ratings is relatively scarce in game recommendation scenarios \cite{Proposal_of_a_Game_Recommendation_System_Considering_Playing_Time_and_Friendships}. Consequently, implicit feedback, represented by playtime, becomes a key indicator of user interest. However, as shown in Fig.\ref{fig_intro}, the playtime data exhibits peculiar characteristics, following a power-law distribution with a significant share of zero values. These zero values may stem from users either lacking interest in the game or acquiring it but postponing engagement. We are the first to investigate the inherent imbalanced nature of implicit feedback in games and emphasize that adeptly addressing this imbalanced feedback is critical for effectively quantifying user preferences in recommendations and uncovering their latent interests.


As per our current understanding, prevailing {category-diversified} recommendation methodologies have only concentrated on scenarios where a one-to-one mapping exists between items and categories \cite{Choosing-the-Best-of-Both-Worlds:-Diverse-and-Novel-Recommendations-through-Multi-Objective-Reinforcement-Learning,Disentangled-Representation-for-Diversified-Recommendations}, where each item belongs to only one category. Nonetheless, in game dataset characterized by extensive cross-category attributes, a single game may simultaneously belong to multiple categories. Notably, exemplified by games like ``Left 4 Dead 2", falling into diverse categories such as ``Action", ``Zombies", ``Co-op", and ``FPS", among others. This necessitates the development of novel approaches to accommodate this one-to-many item-category mapping, posing a significant challenge to achieve diversity by covering a wider array of categories within the recommended game list. 


Recent advancements in Graph Neural Networks (GNNs) have introduced a diversified recommendation approach, 
facilitating the generation of embeddings
that encompass a broader spectrum of category information through controlled neighbor aggregation for user nodes \cite{DGCN,DGRec}. 
However, the existing methods exhibit limitations when extended to the domain of game recommendation. Primarily, the imbalanced distribution of implicit feedback imparts a bias to the model, favoring the recommendation of games akin to those with long gaming durations that the user has engaged with. 
As a consequence, this bias results in overlooking a considerable number of items with low durations or zero playtimes situated within the long tail. Notably, these items have indeed been interacted with by the user, harboring latent interests. Furthermore, previous models confront challenges in effectively addressing intricate relationships among multi-category games, rendering them inadequate in detecting redundancies among the neighbors of game/player nodes. Consequently, this deficiency hampers the successful execution of diversified information aggregation strategies.

To effectively tackle the dual challenges delineated above, 
{we present a novel framework, named DRGame, including a two-fold strategy.}
First, we introduce \underline{B}alance-driven \underline{I}mplicit \underline{P}references \underline{L}earning (BIPL), a component aimed at equalizing the distribution of game playtime among different players while concurrently assessing user preferences from a multi-categorical perspective. By adeptly harmonizing these two facets, we establish a robust foundation of reliable data support for subsequent diversified recommendations. 
{Second, we propose \underline{C}lustering-based \underline{D}iversified \underline{R}ecommendation \underline{M}odule (CDRM). It} initiates by extracting category-aware embeddings for players and games. Building on these enriched embeddings, we engage a clustering mechanism, guaranteeing that players or games within the same cluster share categorical similarities.
In this context, representative items are meticulously selected from each cluster for each player or game, grounded in balanced implicit preferences. Additionally, we design a novel method that asymmetrically aggregates the neighbor information of players and games. Lastly, we strategically adjust the weights of different games during the training process to further increase the impact of long-tail items.

Our contributions can be summarized as follows:
\begin{itemize}

\item 
Our work represents the first endeavor to achieve diversified recommendations specifically tailored for video games, and it addresses the unique challenges posed by games with multiple categories.

\item We design a novel approach that seeks to balance implicit preferences by integrating both individual player-game interactions and broader player-category interests, providing essential data support for subsequent efforts in achieving diversified recommendations.


\item We propose a novel diversity-directed asymmetric neighbor aggregation method within graph-based recommendation models, aiming to enhance the representation and exploration of users' underlying diverse interests, ultimately leading to improved recommendation diversity.


\item We conduct a comprehensive evaluation of the effectiveness of our proposed method on a real-world dataset, showcasing the superiority of our proposed method compared to the state-of-the-art approaches in video game recommendation.

\end{itemize}

\section{Related Work}
\subsection{Recommender System for Video Games}
Video game recommender systems have attracted significant attention in academia. Implicit preference has been a key focus in this context. Studies like \cite{From-Implicit-Preferences-to-Ratings:-Video-Games-Recommendation-based-on-Collaborative-Filtering,Proposal_of_a_Game_Recommendation_System_Considering_Playing_Time_and_Friendships} highlight the limitations of explicit feedback data like ratings in the video game field due to their scarcity and quality. As an alternative, they propose utilizing implicit feedback data, such as game playtime, as a representation of user interests. \cite{Hybrid-system-for-video-game-recommendation-based-on-implicit-ratings-and-social-networks} improves the handling of game playtime by employing estimation rating methods similar to those used in the music domain. Additionally, \cite{A-hybrid-recommender-system-for-steam-games} integrates game playtime with other data to calculate ratings. 

Methodologically, most of the mentioned works rely on traditional collaborative filtering approaches. Recent studies like \cite{Recommender-Systems-for-Online-Video-Game-Platforms:-The-Case-of-STEAM} evaluate the performance of several existing models in game recommendation, including metrics such as accuracy and diversity.
Furthermore, \cite{SCGRec} employs a graph-based recommendation model that incorporates game playtime, social network, and interaction data. As the game industry evolves and subscription service models gain popularity, the significance of recommendation diversity becomes more pronounced.

In the existing literature, diversity has not been given substantial attention in this particular context. Therefore, our work places significant emphasis on addressing this aspect.

\subsection{Diversified Recommendation}

Since the initial introduction of diversified recommendation by \cite{Improving-Recommendation-Lists-through-Topic-Diversification}, a range of algorithms has emerged to address this particular challenge. Notably, methods based on Maximum Marginal Relevance (MMR) \cite{MMR,Managing-Diversity-in-Airbnb-Search,Towards-Results-Level-Proportionality-for-Multi-Objective-Recommender-Systems,Feature-aware-diversified-re-ranking-with-disentangled-representations-for-relevant-recommendation} and Determinantal Point Process (DPP) \cite{DPP,Diversified-Interactive-Recommendation-with-Implicit-Feedback,Enhancing-Recommendation-Diversity-Using-Determinantal-Point-Processes-on-Knowledge-Graphs,Sliding-Spectrum-Decomposition-for-Diversified-Recommendation} have been proposed, which directly reorders the output results of the recommendation model to enhance diversity. 



\begin{figure*}[t]
\vspace{-3mm}
\centering
\includegraphics[width=1\textwidth]{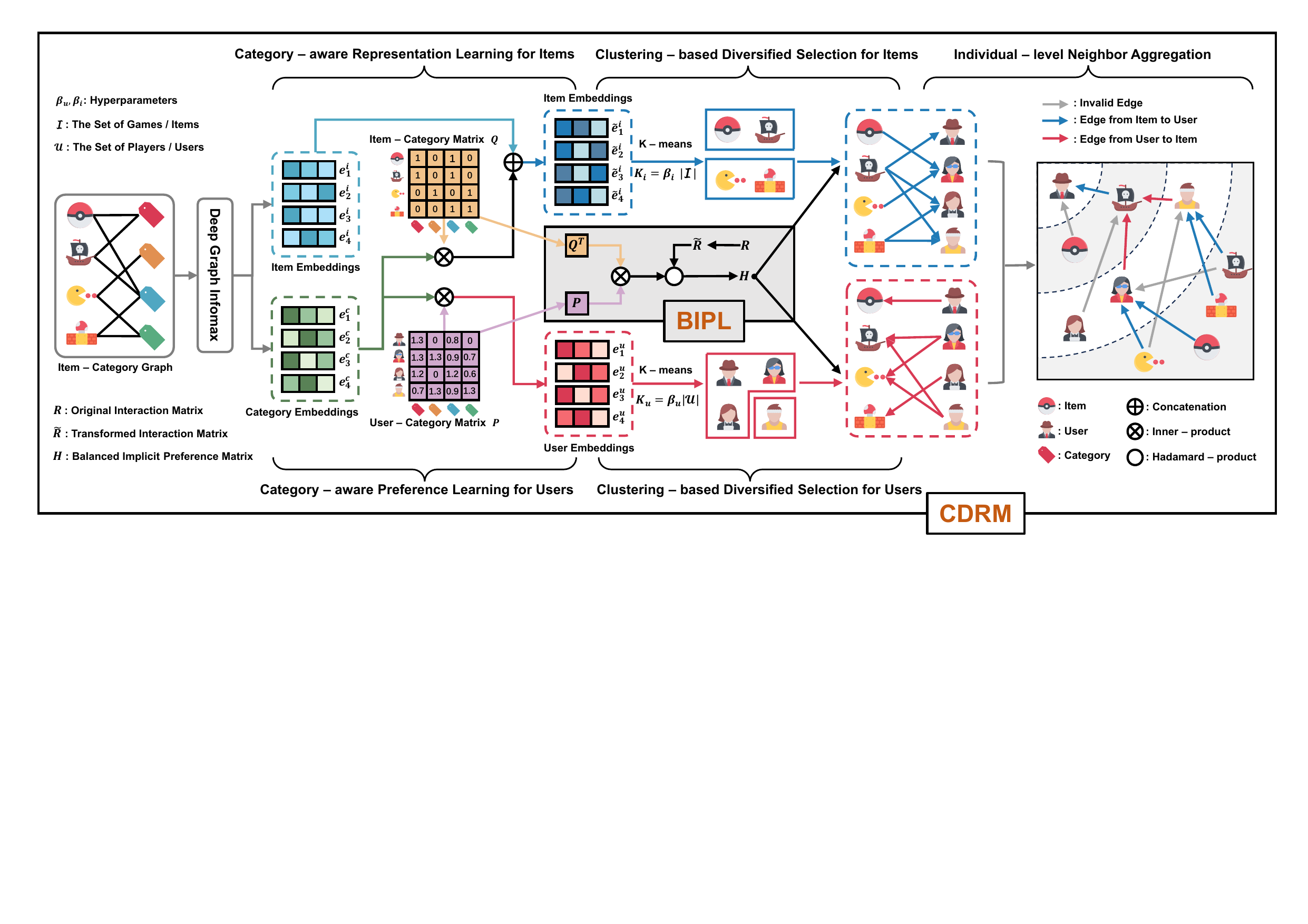} 
\vspace{-61mm}
\caption{Illustration of our proposed DRGame framework. The center of the graphical framework is BIPL: Balance-driven Implicit Preferences Learning. The surrounding segments constitute CDRM: Clustering-based Diversified Recommendation Module, consisting of three sequential constituents, namely Category-aware Representation Learning, Clustering-based Diversified Node Selection, and Individual-level Neighbor Aggregation, proceeding from left to right.}
\label{fig_overview} \vspace{-6mm}
\end{figure*}

Recently, GNNs have gained popularity in recommendation systems \cite{tian2022reciperec}, offering a fresh perspective on enhancing diversity through the exploration and utilization of high-order neighbors in graph structures. To the best of our knowledge, DGCN \cite{DGCN} was the first work to integrate GNN-based recommendation models with diversity strategies. 
It adjusts sampling probability by reversing the histogram of category statistics for neighboring items. Another noteworthy work, DGRec \cite{DGRec}, introduces Submodular Neighbor Selection during the embedding aggregation process. Additionally, RGCF \cite{RGCF} focuses on denoising to improve accuracy while maintaining diversity.

In our DRGame, we incorporate an optimized neighbor aggregation strategy similar to previous models. However, unlike existing approaches, our method takes into account both the user and item node embedding generation processes and effectively handle the unique scenario of multiple categories associated with individual games. 


\section{Methodology}
\subsection{Problem Formulation}
In this study, we consider a set of players or users denoted as $\mathcal{U} = \{u_1, u_2, \dots, u_{\left | \mathcal{U} \right |}\}$ and a set of games or items denoted as $\mathcal{I} = \{i_1, i_2, \dots, i_{\left | \mathcal{I} \right |}\}$. Here, $\left | \mathcal{U} \right |$ and $\left | \mathcal{I} \right |$ are the number of players and games, respectively. Each player $u \in \mathcal{U}$ has a recorded time spent on game $i \in \mathcal{I}$ denoted as $t_{u,i}$, where $t_{u,i} \geq 0$. A value of $t_{u,i}=0$ indicates that player $u$ has acquired game $i$, but has not yet engaged with it. To represent the observed player-game interactions, we use a matrix $\boldsymbol{R} \in \mathbb{R}^{\left | \mathcal{U} \right | \times \left | \mathcal{I} \right |}$, where the element $R_{u,i} = t_{u,i}$ if there exists an interaction between player $u$ and game $i$, otherwise $R_{u,i} = -1$.
To develop a graph neural network (GNN)-based recommender model, we construct a bipartite graph $\mathcal{G}=(\mathcal{V},\mathcal{E})$ based on the player-game interaction data $\boldsymbol{R}$. Here, $\mathcal{V}=\mathcal{U} \cup \mathcal{I}$, meaning that $\mathcal{V}$ includes both player nodes and game nodes. An edge $e_{u,i} \in \mathcal{E}$ between the player node $u$ and game node $i$ exists in the graph $\mathcal{G}$ if $R_{u,i} \geq 0$. 

In this context, we also have a category set $\mathcal{C} = \{c_1, c_2, \dots, c_{\left | \mathcal{C} \right |}\}$ associated with the game set $\mathcal{I}$. The category data for games is represented by a binary matrix $\boldsymbol{Q} \in \mathbb{R}^{\left | \mathcal{I} \right | \times \left | \mathcal{C} \right |}$, where $Q_{i,c} = 1$ if game $i$ belongs to category $c$, otherwise $Q_{i,c} = 0$.


The main goal of this research is to design a recommender system that offers personalized game recommendations to player $u$. The system suggests a top-$N$ list of games $\{i_1, i_2, \dots, i_N\}$ that player $u$ has not previously interacted with. The diversified recommendation is to ensure that the recommended games cover a wider range of categories from $\mathcal{C}$. Thus, the proposed recommender system aims to deliver a diverse selection of games to player $u$, offering a comprehensive gaming experience across various game categories.

\vspace{-2mm}
\subsection{Overview}
As Fig.\ref{fig_overview} depicts, the proposed framework DRGame consists of two essential components, namely Balance-driven Implicit Preferences Learning (BIPL) for data pre-processing and Clustering-based Diversified Recommendation Module (CDRM) for final prediction.

BIPL addresses the challenge of imbalanced implicit feedback in game interactions and incorporates diverse user interest preferences across distinct categories, providing crucial data support for subsequent modeling.

CDRM comprises four consecutive stages: Category-aware Representation Learning, Clustering-based Diversified Node Selection, Individual-level Neighbor Aggregation, and Model Training. Category-aware Representation Learning focuses on representing items and users within the various categories of the game domain. Clustering-based Diversified Node Selection aims to eliminate redundant similar nodes in the interaction bipartite graph, leading to more diverse node embeddings for the following Individual-level Neighbor Aggregation. Lastly, by adjusting the loss weights for different games in Model Training, we further enhance the exposure of disadvantageous categories. 

The integration of these components results in the realization of a diversified recommendation system tailored for video games.

\vspace{-2mm}
\subsection{The Proposed Data Pre-processing: BIPL}
Given the matrix $\boldsymbol{R}$ representing player-game interactions and denoting the actual playing time duration, it is evident that users' average time spent on different games varies significantly. Consequently, directly using this matrix to assess player preferences is deemed inadequate. Moreover, from a global perspective, the distribution of playing time follows a power-law distribution, leading to a highly imbalanced pattern where a few dominant items overshadow numerous disadvantaged items. Inspired by the approach in \cite{SCGRec}, we propose a transformation of the element values in matrix $\boldsymbol{R}$ while preserving its structural integrity:
\begin{equation}
\tilde{R}_{u,i}=\frac{\# (u' | 0 \le R_{u',i} \le R_{u,i} )}{\#(u' | R_{u',i} \ge 0)}, \label{XX}
\end{equation}
where $\boldsymbol{\tilde{R}}$ denotes the transformed interaction matrix. In $\boldsymbol{\tilde{R}}$, the term $\#(u' | R_{u',i} \ge 0)$ represents the number of players who have interacted with game $i$, including those players whose playing time is $0$. Similarly, the term $\# (u' | 0 \le R_{u',i} \le R_{u,i} )$ calculates the number of users in game $i$ whose time investment is less than or equal to that of user $u$. Incorporating this transformation, the original matrix $\boldsymbol{R}$ undergoes a transition from -1 to 0 for non-existent interaction pairs, and the values for interaction pairs with a playing time of 0 in $\boldsymbol{R}$ are set to values greater than 0 in the transformed matrix $\boldsymbol{\tilde{R}}$. This transformation addresses the imbalance in player interaction in terms of playing times, effectively relieving the associated issues. As a result, the distribution of interactive ratings in $\boldsymbol{\tilde{R}}$ exhibits a departure from the long-tail distribution, effectively mitigating the imbalance between high and low gaming durations.

\vspace{-4mm}
\begin{figure}[H]
    \centering
    \subfigure[$\boldsymbol{\tilde{R}Q}$]{
        \includegraphics[width=1.54in]{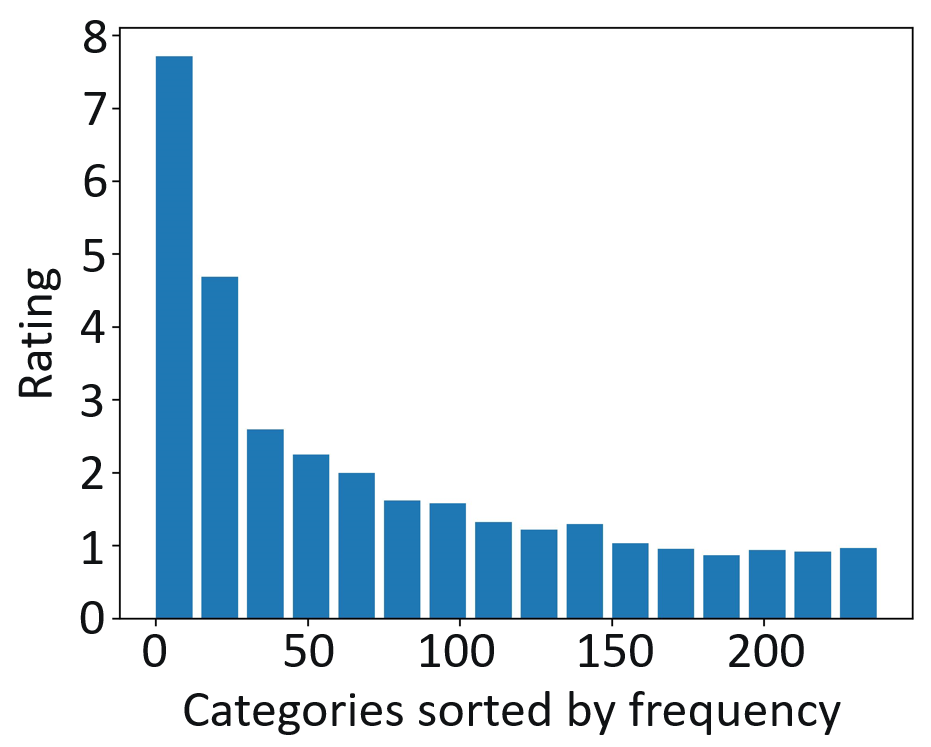}
    }
    \subfigure[$\boldsymbol{P}$]{
	\includegraphics[width=1.59in]{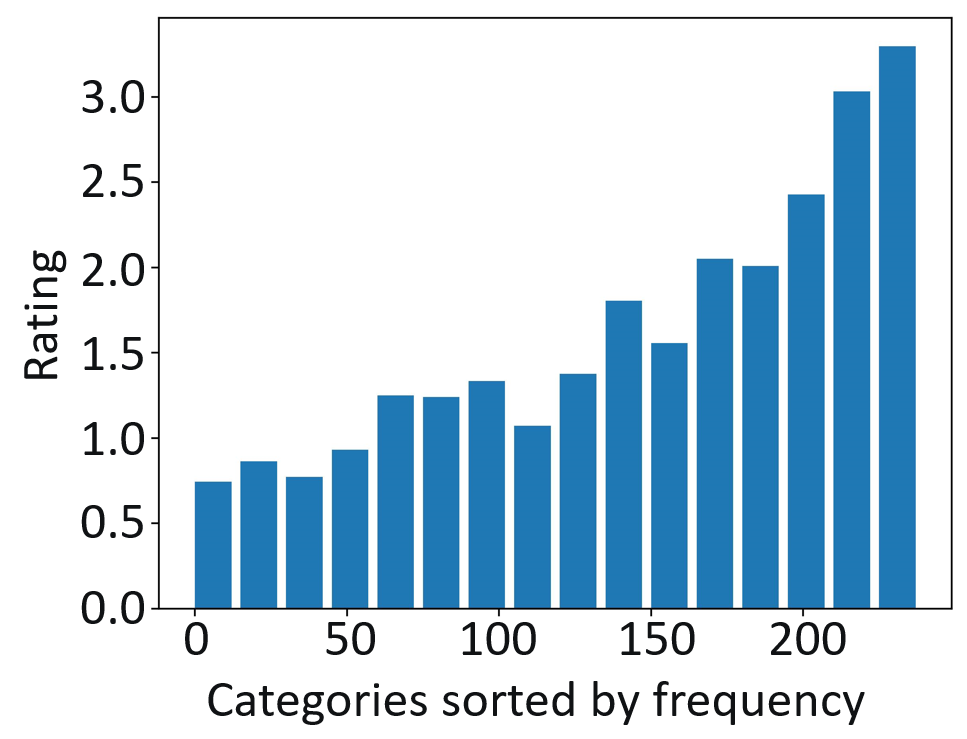}
    }\vspace{-4mm}
    \caption{(a) The original ratings provided by players towards different categories; (b) The processed ratings provided by players towards different categories. The processed ratings increases the exposure of low frequency categories.
}
    \label{mat_P}
    \vspace{-4mm}
\end{figure}

To capture the diverse levels of player interest in different game categories and to better explore the potential games that players may engage with, we introduce a user-category matrix $\boldsymbol{P} \in \mathbb{R} ^ {\left | \mathcal{U} \right | \times \left | \mathcal{C} \right | }$ to represent player preferences towards categories. The elements $P_{u,c}$ in matrix $\boldsymbol{P}$ are computed as:
\begin{equation}
P_{u,c}=\frac{p^{local}_{u,c}}{p^{global}_{u,c}}, \label{XX}
\end{equation}
where
\begin{equation}
\begin{split}
& p^{local}_{u,c} = \frac{\# (i|Q_{i,c}=1 \wedge R_{u,i}\ge 0)}{\# (i|R_{u,i}\ge 0)} \\
& p^{global}_{u,c} = \frac{\# (i|Q_{i,c}=1)}{\# (i|Q_{i,c'}=1 \wedge (\tilde{R}Q)_{u,c'} > 0)}
\label{XX}
\end{split},
\end{equation}
where $p^{local}_{u,c}$ denotes the ratio of games that player $u$ has interacted with, which are assigned the category $c$, out of the total interactions of this user. And $p^{global}_{u,c}$ indicates the ratio of games with the category $c$ to the games with categories that match the categories of all items that player $u$ has interacted with, regardless of whether player $u$ has interacted with those games. Indeed, directly using $\boldsymbol{\tilde{R}Q}$ to represent $\boldsymbol{P}$ is a viable option, but it may lead to certain high-frequency category overshadowing the player's true interests, especially when the distribution of different category within the game collection $\mathcal{I}$ is unbalanced. 
As depicted in Fig.\ref{mat_P}(a), the computation approach of $\boldsymbol{\tilde{R}Q}$ reveals a positive correlation between users' preference ratings for a category and the frequency of that category. Higher frequency implies higher preferences.
To address these potential biases and provide a more accurate representation of player preferences towards different game categories, while simultaneously increasing the exposure of disadvantaged categories, we propose an alternative approach. Specifically, we define $P_{u,c}$ as the ratio between the local proportion of category $c$ within the player's interactions and its global proportion across the entire game collection $\mathcal{I}$. This normalization ensures that the player-category matrix $\boldsymbol{P}$ takes into account both the player's individual interests and the overall distribution of game categories in the dataset, resulting in a more balanced and comprehensive view of player preferences, as shown in Fig.\ref{mat_P}(b).

\begin{figure}[H]
\vspace{-4mm}
    \centering
    \subfigure[$\boldsymbol{R}$]{
        \includegraphics[width=1.54in]{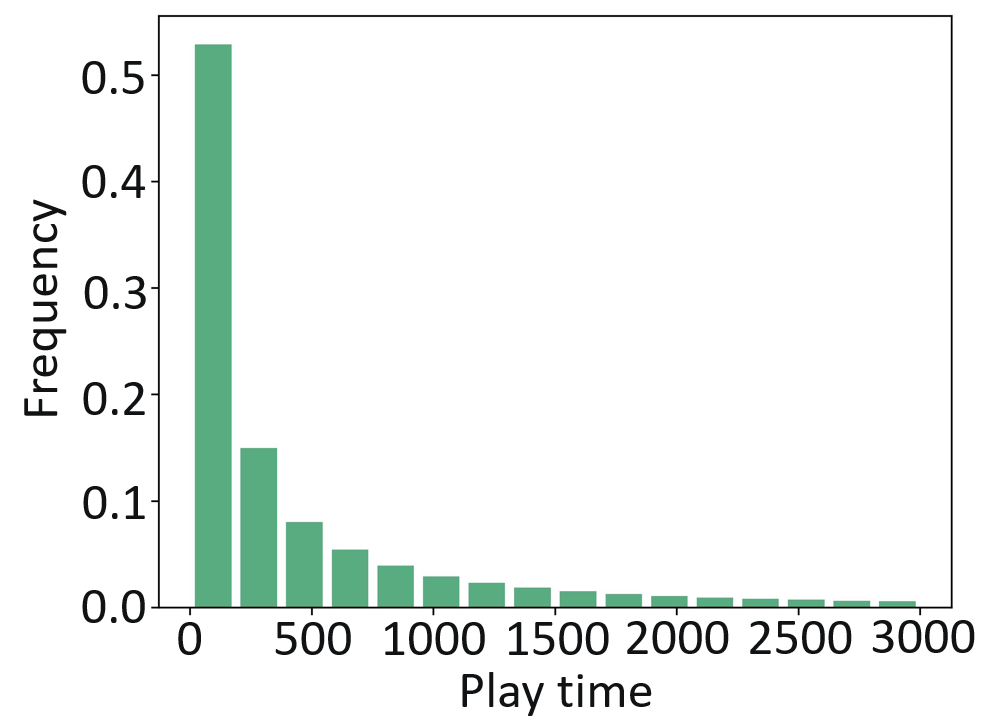}
    }
    \subfigure[$\boldsymbol{H}$]{
	\includegraphics[width=1.58in]{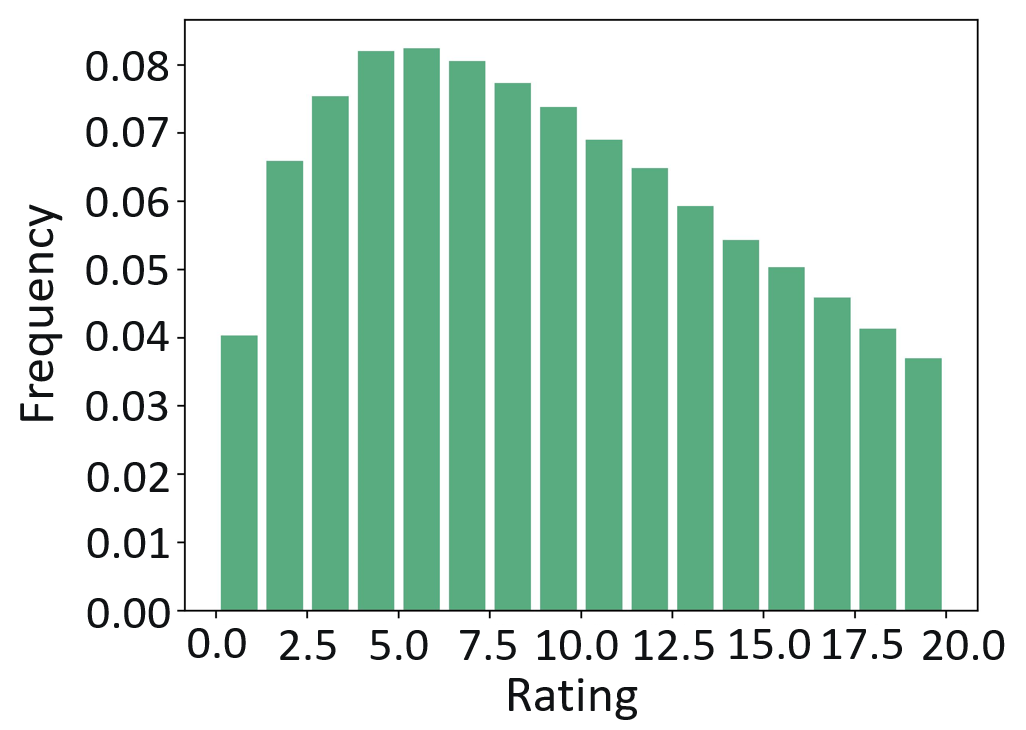}
    }\vspace{-4mm}
    \caption{The distribution of elements in matrix (a) $\boldsymbol{R}$ and matrix (b) $\boldsymbol{H}$. }
    \label{mat_R_H}
    \vspace{-4mm}
\end{figure}

Combining the discussed player's game interactions and category interests, the balanced implicit preferences for video games can be calculated as  follows:
\begin{equation}
\boldsymbol{H} = \boldsymbol{PQ^{T}} \circ \boldsymbol{\tilde{R}}, \label{XX}
\end{equation}
where $\boldsymbol{H} \in \mathbb{R} ^ {\left | \mathcal{U}  \right |  \times \left | \mathcal{I}  \right | }$ and $\circ$ is the Hadamard product. If player $u$ has interacted with game $i$ but has never played it (playing time is zero), $H_{u,i}$ still takes into account the player's preferences, which can vary based on the category interests of this user.
For games that player $u$ has never interacted with, $H_{u,i}$ is assigned a rating of 0, indicating no previous preference for these unexplored games. The implicit preferences encoded in $H$ are essential in delivering personalized and varied game suggestions by considering both the individual player-game interactions and broader player-category interests. As illustrated in Fig.\ref{mat_R_H}(a), the distribution of elements in the matrix $\boldsymbol{R}$, which represent the gaming time, initially exhibits an imbalanced long-tail distribution from the implicit feedback. However, our data pre-processing method successfully transforms this distribution into a more uniformly distributed rating form, as demonstrated in Fig.\ref{mat_R_H}(b).

\subsection{The Proposed Recommendation Model: CDRM}
\subsubsection{Category-aware Representation Learning.}
To explore similarities among games, we transform the matrix $\boldsymbol{Q}$ into an undirected graph and apply Deep Graph Infomax (DGI) \cite{DGI} on it. DGI, being an unsupervised method specifically designed for learning node representations in graph-structured data, 
proves to be well-suited for uncovering intricate relationships among games with multiple categories. By leveraging DGI, we obtain two outputs: item embeddings $\boldsymbol{E_{I}} \in \mathbb{R} ^ {\left | \mathcal{I}  \right |  \times d }$ and category embeddings $\boldsymbol{E_{C}} \in \mathbb{R} ^ {\left |\mathcal{C}  \right |  \times d }$, where $d$ is the dimension of the embeddings.

In order to distinguish the preference patterns of different player groups with respect to categories and eliminate redundant similarity among user nodes during the subsequent node selection process, we propose an approach to obtain category-aware representation embeddings for users by weighting category embeddings in $E_{C}$ with user interests towards categories in $\boldsymbol{P}$, as shown below: 
\begin{equation}
\begin{split}
 \boldsymbol{E_{U}}= \boldsymbol{P} \otimes \boldsymbol{E_{C}}, 
\label{XX}
\end{split}
\end{equation}
where $\boldsymbol{E_{U}} \in \mathbb{R} ^ {\left | \mathcal{U}  \right |  \times d }$ and $\otimes$ is the matrix inner-product. Similarly, to refine item embeddings by considering features across different categories, we use the following technique:
\begin{equation}
\begin{split}
 \boldsymbol{\tilde{E}_{I}}= (\boldsymbol{Q} \otimes \boldsymbol{E_{C}} )\oplus \boldsymbol{E_{I}},
\label{XX}
\end{split}
\end{equation}
where $\boldsymbol{\tilde{E}_{I}} \in \mathbb{R} ^ {\left | \mathcal{I}  \right |  \times 2d }$ and $\oplus$ is the matrix concatenation. 

The computation process of $\boldsymbol{E_{U}}$ and $\boldsymbol{\tilde{E}_{I}}$ can be observed in Fig.\ref{fig_overview}, where $\boldsymbol{E_{U}}$ denotes the preference of players for categories and $\boldsymbol{\tilde{E}_{I}}$ denotes the features of games based on those categories.

\subsubsection{Clustering-based Diversified Node Selection.}
In the user and item sets,
it is evident that certain players share similar preferences at the category level, and likewise, some games exhibit similarities in their themes. However, to ensure that each node embedding in the subsequent graph-based recommendation model accurately captures a diverse array of neighbor information, it is imperative to mitigate the potential influence of numerous similar neighbors on the embedding learning process. To achieve this, we conduct node clustering with a focus on category-aware user and item representations. Specifically, we perform \textit{k}-means clustering individually on each set. For players, we set the number of clusters $k_{u}$ to $ \beta_u\left | \mathcal{ \mathcal{U} } \right | $, and for games, we set $k_{i}$ to $ \beta_i \left | \mathcal{ \mathcal{I} } \right |$. 
Following these clustering operations, the player and game set can be respectively denoted as $\mathcal{U} =\{U_1, U_2,\dots,U_{k_{u}}\}$ and $\mathcal{I} =\{I_1, I_2,\dots,I_{k_{i}}\}$. To eliminate redundancy in interactions, we adopt a node selection approach for each player $u$ and each game $i$ in graph $\mathcal{G}$. This approach involves selecting only the most representative nodes from each cluster of players and games. Specifically, for each player $u$, we choose the game node with the highest implicit preference rating as the representative node for that cluster, following the equation:
\begin{equation}
\begin{split}
& \mathcal{E}_{u}^{I\to U}=\bigcup_{j=1}^{k_i} (u, \underset{i\in I_j}{\mathrm{argmax}} \ H_{u,i}), 
\label{node_selection_u}
\end{split}
\end{equation}
where $H_{u,i}$ represents the implicit preference rating of player $u$ towards game $i$. The set $\mathcal{E}_{u}^{I\to U}$ comprises edges that connect game nodes to player nodes, with player $u$ being the destination of these edges. The operation $\mathrm{argmax}_{i\in I_j} \ H_{u,i}$ selects the game from the cluster $I_j$ that has the highest implicit preference rating for user $u$. 

Similarly, for each game $i$, we also select the game node with the highest implicit preference rating as the representative node for that cluster, as given by the equation:

\vspace{-4mm}
\begin{equation}
\begin{split}
& \mathcal{E}_{i}^{U\to I}=\bigcup_{j=1}^{k_u} (\underset{u\in U_j}{\mathrm{argmax}} \ H_{u,i},i). 
\label{node_selection_i}
\end{split}
\end{equation}
\vspace{-2mm}

According to the two types of edge sets obtained from the above node selection approach, we can decompose the bipartite interaction graph $\mathcal{G}$ into two subgraphs:
\begin{equation}
\begin{split}
 \mathcal{G}^{I\to U}=(\mathcal{V},\bigcup_{u\in \mathcal{U}} \mathcal{E}_{u}^{I\to U}), \mathcal{G}^{U\to I}=(\mathcal{V},\bigcup_{i\in \mathcal{I}} \mathcal{E}_{i}^{U\to I}).
\label{xx}
\end{split}
\end{equation}
\vspace{-2mm}

The process of clustering and node selection is visualized in Fig.\ref{fig_overview}, demonstrating its significance in preparing the data for efficient neighbor aggregation and enhanced recommendation performance in the subsequent steps.

\subsubsection{Individual-level Neighbor Aggregation.}

Upon obtaining the directed graph $\mathcal{G}^{I\to U}$, which includes edges from game to player nodes, and the directed graph $\mathcal{G}^{U\to I}$, which includes edges from player to game nodes, we proceed to employ a GNN with a message passing scheme to learn embeddings of users and items. Specifically, we utilize LightGCN \cite{LightGCN} as the backbone GNN layer. LightGCN is a concise and efficient graph-based recommendation model that forgoes the use of feature transformation and nonlinear activation, employing simple weighted sum aggregators. In contrast to its approach of aggregating all neighbor information when learning node representations, we introduce an Individual-level Neighbor Aggregation strategy. In this strategy, we asymmetrically aggregate the neighbor embeddings of players and game nodes separately as follows:
\begin{equation}
\begin{split}
& \boldsymbol{e} ^{(l+1)}_{u}=\sum_{i \in \mathcal{N}_{u}}\frac{\mathbb{I}((u,i)\in \mathcal{G}^{I\to U})}{\sqrt{|\mathcal{N}_{u}|}\sqrt{|\mathcal{N}_{i}|}} \boldsymbol{e} ^{(l)}_{i} \\
& \boldsymbol{e} ^{(l+1)}_{i}=\sum_{u \in \mathcal{N}_{i}}\frac{\mathbb{I}((u,i)\in \mathcal{G}^{U\to I})}{\sqrt{|\mathcal{N}_{i}|}\sqrt{|\mathcal{N}_{u}|}} \boldsymbol{e} ^{(l)}_{u}
\label{Neighbor_Aggregation}
\end{split},
\end{equation}
where $\boldsymbol{e} ^{(l)}_{u}$ and $\boldsymbol{e} ^{(l)}_{i}$ respectively denote the embeddings of user $u$ and item $i$ after $l$ layers of propagation. The function $\mathbb{I}(\cdot )$ is a binary indicator returning 1 when the condition is true. $\mathcal{N}_{u}$ and $\mathcal{N}_{i}$ denote the neighbor set of user $u$ and item $i$ in the graph $\mathcal{G}$, respectively. The normalization term $1/(\sqrt{|\mathcal{N}_{u}|}\sqrt{|\mathcal{N}_{i}|})$ is generated based on the original graph $\mathcal{G}$ rather than the node-selected graphs $\mathcal{G}^{I\to U}$ and $\mathcal{G}^{U\to I}$ for consistency with the subsequent model training process.

The visualization of asymmetric aggregation can be seen in Fig.\ref{fig_overview}. Compared to the symmetric aggregation based on user-item interactions, our approach allows for the selection of the most representative neighbors specifically tailored to individual players or games. This selection disregards redundant items with higher similarity, ensuring the diversity of embedding representations.

After $L$ layers according to Eq. \ref{Neighbor_Aggregation}, we further combine the embeddings obtained at each layer to form the final representation, donated as $\boldsymbol{e} _{u}$ and $\boldsymbol{e} _{i}$:
\begin{equation}
\begin{split}
\boldsymbol{e}_{u}=\sum_{l=0}^{L} \alpha_l \boldsymbol{e} ^{(l)}_{u}, \ \  \boldsymbol{e}_{i}=\sum_{l=0}^{L} \alpha_l \boldsymbol{e} ^{(l)}_{i}, 
\label{XX}
\end{split}
\end{equation}
where $\alpha_l$ is a hyper-parameter to denote the importance of $l$-th layer embedding. It can be manually adjustable, and here we set it as the output of an attention network \cite{Attention} to be optimized automatically.

Finally, we employ the inner product of player and game embeddings to predict the likelihood of interaction between them:

\vspace{-6mm}
\begin{equation}
\begin{split}
\hat{y}_{u,i}= \boldsymbol{e}_u\cdot \boldsymbol{e}_i,
\label{XX}
\end{split}
\end{equation}
where $\hat{y}_{u,i}$ is the predicted rank score for recommendation.

\subsubsection{Model Training.}

For the model training process, we address the issue of imbalanced user interactions in the graph $\mathcal{G}$ by randomly sampling negative instances for each observed actual interaction. This approach allows us to construct pairwise training data. To optimize the training, we adopt the Bayesian Personalized Ranking (BPR) loss \cite{BPR}, which is defined as follows:
\begin{equation}\resizebox{0.85\hsize}{!}{ $
\begin{split}
\mathcal{L}=-\sum_{u \in \mathcal{U}} \sum_{i \in \mathcal{N}_u} \sum_{j \in \mathcal{I}, j \not{\in}\mathcal{N}_u}\boldsymbol{w}_i \log \sigma (\hat{y}_{u,i}-\hat{y}_{u,j})+\lambda {\left \| \Theta  \right \|}_{2}^{2},
\label{XX}
\end{split}
$}
\end{equation}
where $\lambda$ controls the strength of $L_2$ regularization and $\sigma(\cdot)$ denotes the sigmoid function. To account for games with disadvantaged categories and enhance their representation during training, we introduce a weight $\boldsymbol{w}_i$ for each game $i$ in the sample loss. The calculation process of $\boldsymbol{w} \in \mathbb{R}^{|\mathcal{I}| \times 1 } $ for all games involves two steps. In the first step, we calculate the number of occurrences of each category in the entire game set $\mathcal{I}$, which is represented as $\boldsymbol{n} \in \mathbb{R}^{|\mathcal{C}| \times 1}$. To exploit categories with low frequency, we obtain $\boldsymbol{\tilde{n}}\in \mathbb{R}^{|\mathcal{C}| \times 1}$ by taking the reciprocal of the elements in $\boldsymbol{n}$. In the second step, we calculate $\boldsymbol{w}$ to assign higher weights to games with low-frequency categories to enhance diversity in recommendations by encompassing a broader range of categories. This two-step procedure can be realized as follows: 
\begin{equation}
\begin{split}
\boldsymbol{n}=\boldsymbol{Q}^T\boldsymbol{I}_{|\mathcal{I}|,1}, \  \boldsymbol{\tilde{n}}=\frac{1}{\boldsymbol{n}} 
\label{XX}
\end{split},
\end{equation}
\begin{equation}
\begin{split}
\boldsymbol{w}=\boldsymbol{Q}(\frac{\boldsymbol{\tilde{n}}}{\left \| \boldsymbol{\tilde{n}} \right \|_1}|\mathcal{C}|)
\label{XX}
\end{split},
\end{equation}
where $\boldsymbol{I}_{|\mathcal{I}|,1}\in \mathbb{R}^{|\mathcal{I}| \times 1}$ represents an all-one vector.

\section{Experiments and Discussion}
\subsection{Experimental Setup}
\subsubsection{Dataset.} 
Adhering to the approach outlined in \cite{SCGRec}, our dataset is derived from the Steam\footnote{https://store.steampowered.com/}. 
It is widely acknowledged as the largest digital distribution platform for video games, extensively utilized in
in the realm of game recommendation research.

To assess the effectiveness of DRGame, we employed the Steam Video Game and Bundle Data shared by \cite{8594844,10.1145/3240323.3240369,10.1145/3077136.3080724}. To ensure data quality, we pre-processed the raw data by removing games with missing category information. Subsequently, we applied the 5-core settings to filter out inactive entities. The statistical details of the filtered dataset are presented in Table \ref{table_dataset}. Finally, we randomly partitioned the interactions into training, validation, and testing sets in a ratio of 6:2:2.

\begin{table}[htbp]
\vspace{-2mm}
\centering
\caption{Statistical Details of the Filtered Dataset.}
\vspace{-2mm}
\begin{tabular}{lr}
\toprule
\midrule
    & \\[-8pt]
    \# Players & 60,742 \\
    \# Games & 7,726 \\
    \# Interactions & 4,145,359 \\
    \# Categories & 315 \\
\bottomrule 
\end{tabular}
\label{table_dataset}
\vspace{-5mm}
\end{table}

\begin{table*}[htbp]
\vspace{-3mm}
\centering
\renewcommand{\arraystretch}{0.75} 
\caption{{Performance Comparision of Different Recommender Systems.}}
\vspace{-3mm}
\begin{tabular}{clccccccccccc}
\toprule

\multirow{2}{*}{}  & \multirow{2}{*}{Method} & \multicolumn{3}{c}{Recall} & \multicolumn{3}{c}{Hit Ratio} & \multicolumn{3}{c}{Coverage} \\
\cmidrule(r){3-5} \cmidrule(r){6-8} \cmidrule(r){9-11}
& & @100 & @150 & @200 & @100 & @150 & @200 & @100 & @150 & @200 \\
\midrule
\multirow{2}{*}{Accuracy-based} & LightGCN & 0.4757 & 0.5402 & 0.6031 & 0.3604 & 0.4280 & 0.4784 & 0.5405 & 0.6535 & 0.7440  \\
\multirow{2}{*}{Methods} & SCGRec & 0.4802 & 0.5437 & 0.5911 & 0.3793 & 0.4299 & 0.4674 & 0.5034 & 0.6211 & 0.7139    \\
& RGCF & \textbf{0.4927} & \textbf{0.5581} & \textbf{0.6049} & \textbf{0.3824} & \textbf{0.4313} & \textbf{0.4788} & 0.5813 & 0.6690 & 0.7482  \\
\midrule
\multirow{3}{*}{Diversity-based} & DDGraph & 0.4065 & 0.4732 & 0.5288 & 0.3059 & 0.3498 & 0.4003 & 0.5856 & 0.6701 & 0.7613  \\
\multirow{3}{*}{Methods} & DGCN       & 0.4033 & 0.4887 & 0.5301 & 0.3043 & 0.3687 & 0.4121 & 0.5901 & 0.6647 & 0.7530  \\
& DGRec      & 0.4149 & 0.4992 & 0.5612 & 0.3187 & 0.3920 & 0.4474 & 0.6377 & 0.7179 & 0.7713  \\
& \textbf{DRGame} & 0.4166 & 0.5070 & 0.5700 & 0.3215 & 0.3983 & 0.4552 & \textbf{0.6824} & \textbf{0.7575} & \textbf{0.8073}  \\

\bottomrule 
\end{tabular} 
\label{tab_comparison} \vspace{-6mm}
\end{table*}

\subsubsection{Metrics.} 
In line with established practices in diversified recommendation \cite{DPP}, we evaluate the effectiveness of our approach through the utilization of top-$N$ metrics, such as Recall and Hit Ratio. In addition, we employ the category Coverage metric to measure diversity. Notably, our dataset contains items that can belong to multiple categories, necessitating consideration of all associated categories when calculating Coverage. Furthermore, the value of $N$ for evaluation is set to 100, 150, and 200, based on the total number of items in the dataset.

\vspace{-1mm}
\subsubsection{Comparison methods.} 
To validate the effectiveness of DRGame, we conducted a comparative analysis against several baselines, including LightGCN \cite{LightGCN}, a widely adopted recommendation model, as well as SCGRec \cite{SCGRec}, a game-specific yet diversity-agnostic approach. We also compared DRGame with RGCF \cite{RGCF}, focusing on denoising while maintaining diversity, and three GNN-based diversified recommendation methods, namely DDGraph \cite{DDGraph}, DGCN \cite{DGCN} and DGRec \cite{DGRec}.


\vspace{-1mm}
\subsubsection{Implementation details.}
In our proposed DRGame, we carefully selected hyperparameters to ensure effective training. Specifically, we set the learning rate to 0.001, the regularization coefficient $\lambda$ to 1e-5, the embedding size to 32, and the batch size to 2048. To prevent overfitting, we implemented an early stopping strategy, whereby training was halted if the performance on the validation set did not improve within 10 epochs. Moreover, it is worth noting that the baseline methods used in the evaluation do not account for the scenario of multiple categories in video game datasets. Therefore, during the testing phase, we randomly selected one of the original multiple categories for each item to maintain consistency in the evaluation process. To comprehensively assess the model's performance, we provide detailed settings of other hyperparameters in the subsequent sections.

\vspace{-2mm}
\subsection{Overall Performance}
\vspace{-1mm}
Table \ref{tab_comparison} presents a comprehensive comparison of DRGame with other baselines. Based on these results, the following observations can be made:
\begin{itemize}
\item
Our proposed DRGame method exhibits superior performance in terms of Coverage, showcasing its leading capabilities in diversifying recommendations.
\item
When considering the accuracy metrics Recall and Hit Ratio, DRGame surpasses other diversity-oriented methods, highlighting its ability to maintain recommendation accuracy while simultaneously enhancing diversity. 
\item
Methods prioritizing recommendation accuracy exhibit higher performance in Recall and Hit Ratio but fall behind DRGame in terms of diversity. 

\item
Despite its integration of a diversity preservation module, RGCF falls short in significantly enhancing diversity. This limitation stems from RGCF's primary objective of enhancing Recall while ensuring no decline in Coverage.


\end{itemize}
Overall, our DRGame method surpasses all diversity-based methods across all metrics in video game recommendation.

\begin{figure}[H]
\vspace{-4mm}
    \centering
    \subfigure[Recall]{
        \includegraphics[width=0.95in]{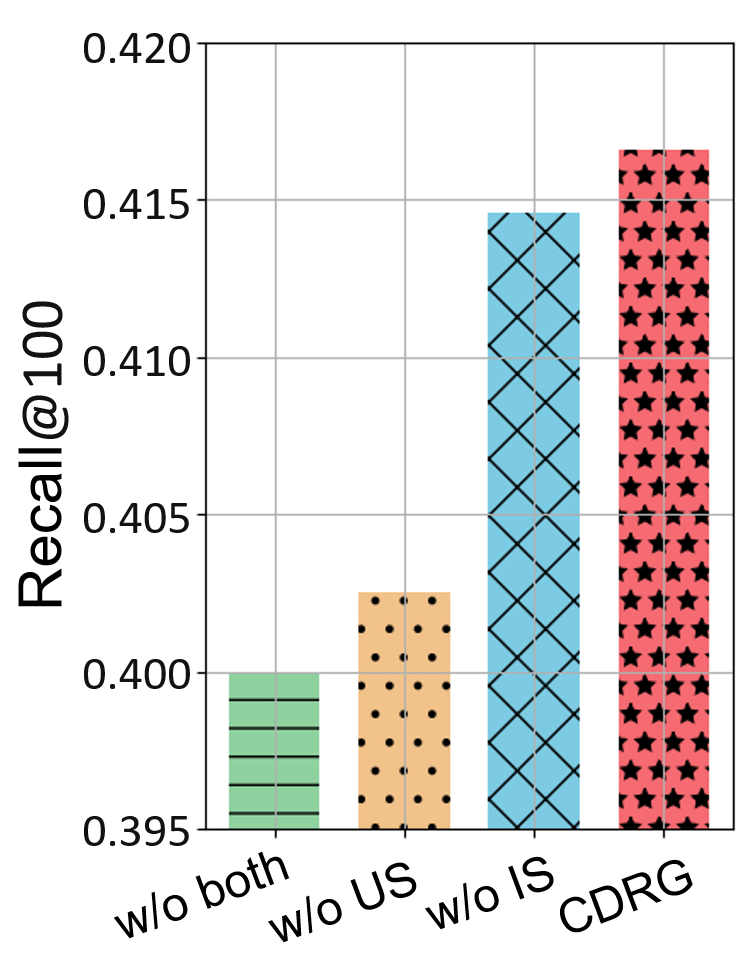}
    }
    \subfigure[Hit Ratio]{
	\includegraphics[width=0.95in]{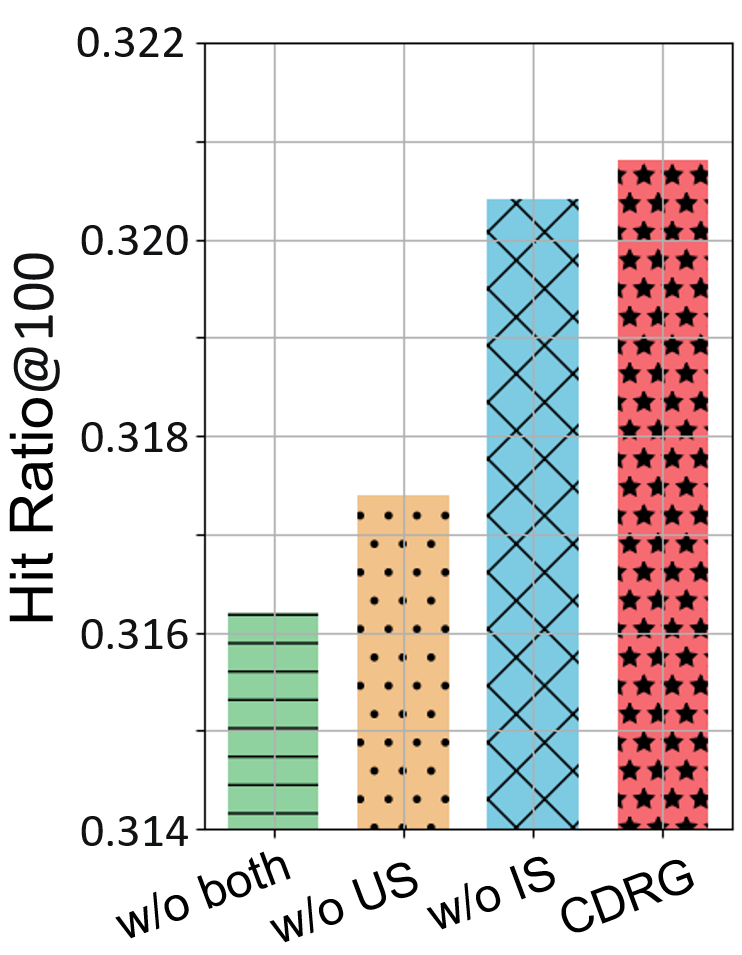}
    }
    \subfigure[Coverage]{
	\includegraphics[width=0.95in]{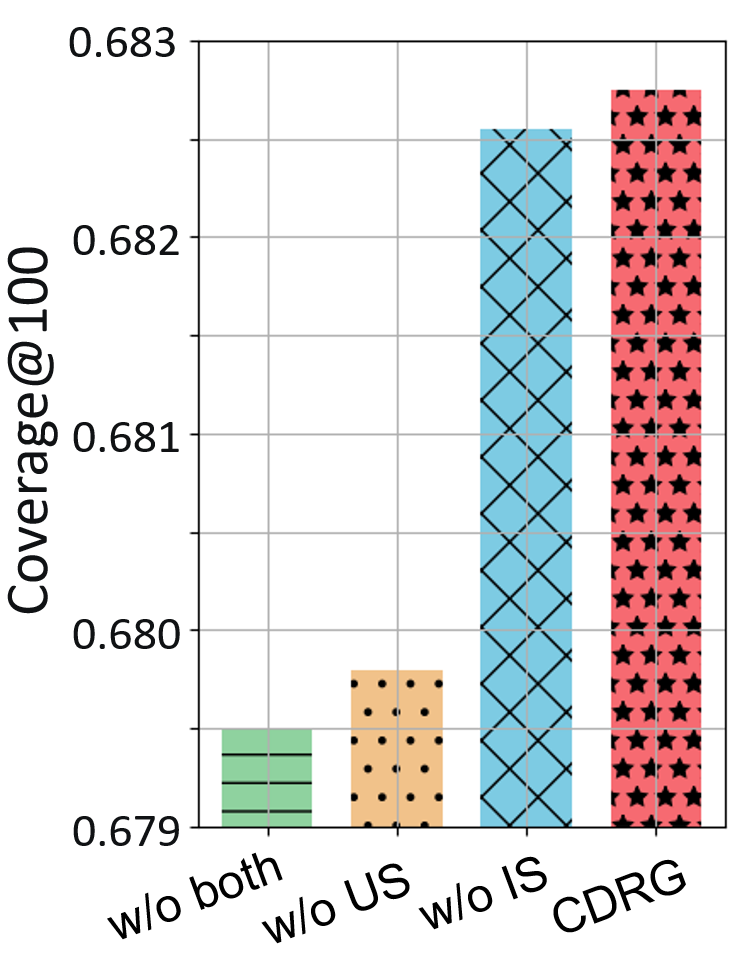}
    }\vspace{-5mm}    
    \caption{Ablation Study on Diversified Node Selection.
}
\label{fig_ablation}
\end{figure}
\vspace{-6mm}

\vspace{-2mm}
\subsection{Model Analyses}
\vspace{-1mm}
\subsubsection{Ablation Study on Diversified Node Selection.}

Ablation experiments investigating diversified node selection are shown in Fig.\ref{fig_ablation}. In these experiments, we explore the consequences of excluding the item selection for users (referred to as ``w/o US"), excluding the user selection for items (referred to as ``w/o IS"), and removing both processes (referred to as ``w/o both"), effectively eliminating the asymmetric neighbor aggregation stage from the framework. The results reveal the significance of user and item selection in enhancing recommendation accuracy and diversity. Notably, the improvement is more pronounced when only item selection for users is performed, compared to only user selection for items. This emphasizes the critical role of eliminating redundant similar item neighbors for users during neighbor aggregation to achieve diversified recommendations.

\vspace{-2mm}
\subsubsection{Sensitivity Analysis on $\beta_i$ and $\beta_u$.}
\begin{figure}[H]
\vspace{-4mm}
    \subfigure[Recall@100]{
        \includegraphics[height=1.25in]{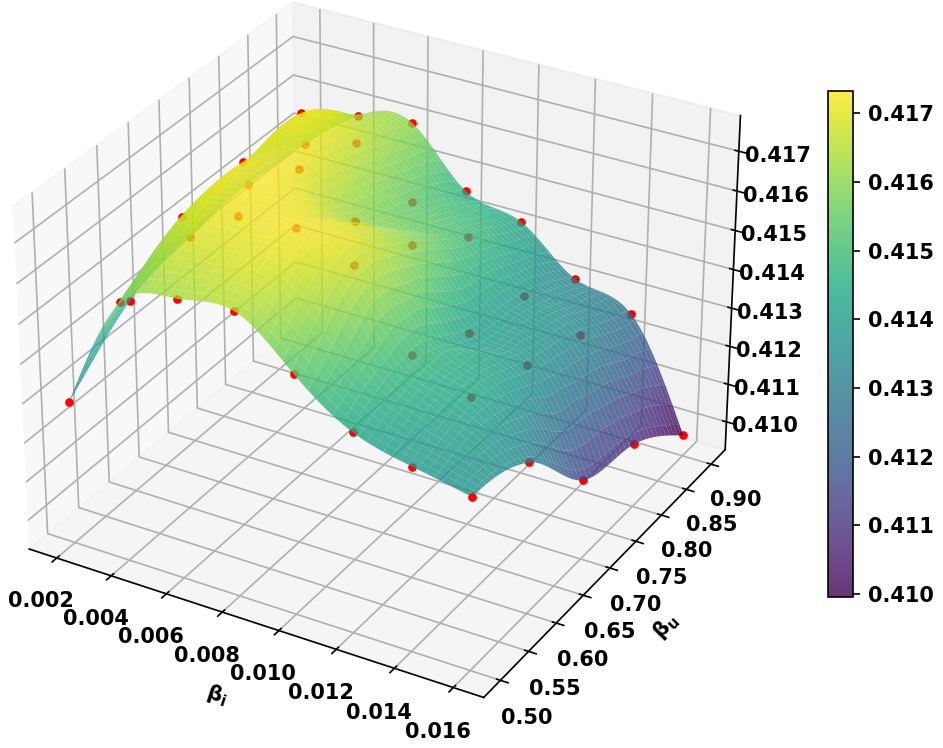}
    }\hspace{0mm}
    \subfigure[Coverage@100]{
	\includegraphics[height=1.25in]{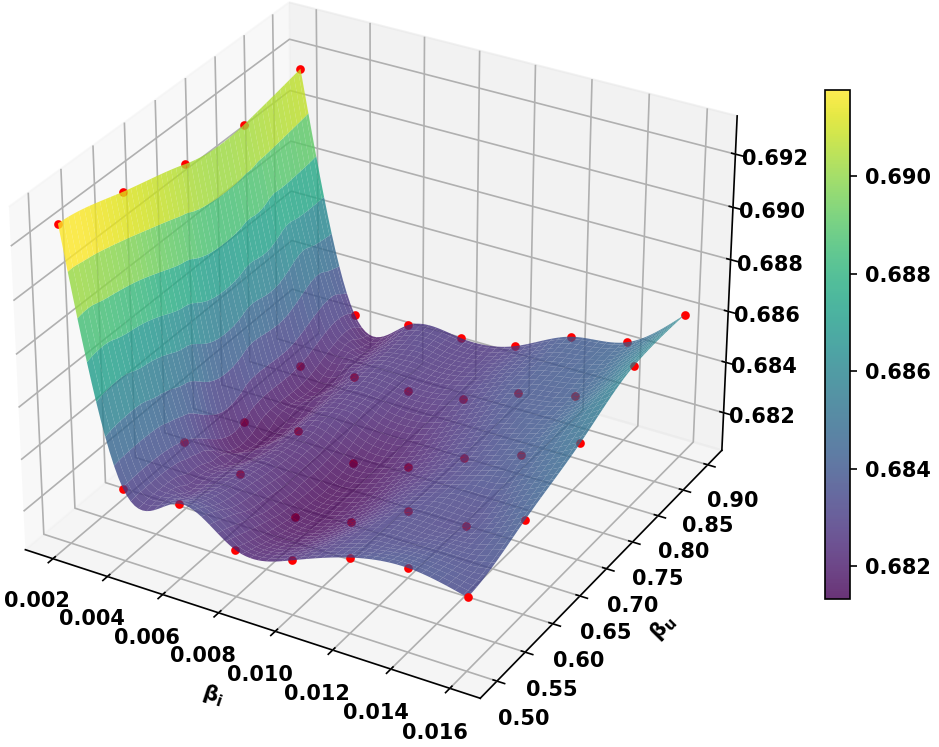}
    }\vspace{-4mm}
    \caption{Sensitivity Analysis on $\beta_i$ and $\beta_u$.}
    \label{fig_sensitivity}
\end{figure}

In Fig.\ref{fig_sensitivity}, we observe the impact of different values for $\beta_i$ and $\beta_u$ on recommendation accuracy and diversity, revealing a trade-off between these two metrics. The X-axis represents the variation of $\beta_i$, where Recall initially experiences a slight increase from 0.02, reaching its peak between 0.04 and 0.08, and then steadily decreasing. Conversely, the trend in Coverage is the opposite, showing a decrease initially and gradually increasing thereafter. The Y-axis in the graph corresponds to the changes in $\beta_u$. Remarkably, when $\beta_u$ is set to 0.6, Recall achieves relatively high performance, while coverage performance lags behind. As our primary focus is enhancing recommendation diversity, we can ensure satisfactory Coverage performance with a lower bound under the same parameter conditions. Consequently, we opt for the parameter setting that optimizes accuracy performance. This analysis provides valuable insights into the sensitivity of our model to various parameter values, aiding in striking an effective balance between recommendation accuracy and diversity.

\vspace{-6mm}




\vspace{-3mm}
\section{Conclusion}
\vspace{-1mm}
In this study, we initiate by introducing BIPL, a novel data pre-processing approach customized for handling implicit feedback in game time, with the goal of achieving a more balanced rating distribution to enable better exploration of users' underlying preferences. Subsequently, we introduce CDRM, a module composed of four consecutive stages, aimed at realizing diversified recommendations. Through evaluations in the context of video game recommendation, we demonstrate that the proposed DRGame, which consists of both BIPL and CDRM, outperforms other diversity-oriented methods within the same domain. Moreover, we conduct ablation experiments to showcase the efficacy of our proposed enhancements to neighbor aggregation in GNNs, resulting in improved diversified recommendations.


\bibliography{aaai24}

\end{document}